\documentclass[preprint,amsmath,amssymb,superscriptaddress]{revtex4-1}
\usepackage{xcolor}
\usepackage{tabu}
\usepackage{titlesec}
\usepackage[utf8]{inputenc}
\usepackage{graphicx,epsfig,epstopdf}
\usepackage{bm}
\usepackage{epstopdf}
\usepackage{amssymb,amsmath,amsfonts,latexsym}
\usepackage{dcolumn}
\usepackage{bm}
\usepackage{hyperref}
\usepackage[utf8]{inputenc}
\newcommand{\nn}{\nonumber\\}

\usepackage{breqn}             
\makeatletter                  
\let\cat@comma@active\@empty   
\makeatother                   
\begin{document}
\title{Prediction of various observables for $B_s^0 \to D_s^{(*)-}\ell^+\nu_\ell$ within covariant confined quark model}%

\author{J. N. Pandya}
\email{jnpandya-phy@spuvvn.edu}
\affiliation{Department of Physics, Sardar Patel University,  Vallabh Vidyanagar 388120, Gujarat,  India.}

\author{P. Santorelli}
\email{pietro.santorelli@na.infn.it}
\affiliation{Dipartimento di Fisica ``E. Pancini”, Universit\`a di Napoli Federico II - Complesso Universitario di Monte S. Angelo Edificio 6, via Cintia, 80126 Napoli, Italy}
\affiliation{INFN sezione di Napoli - Complesso Universitario di Monte S. Angelo Edificio 6, via Cintia,
80126 Napoli, Italy}

\author{N. R. Soni}
\email{nakulphy@gmail.com,  nakul.soni@na.infn.it}
\affiliation{INFN sezione di Napoli - Complesso Universitario di Monte S. Angelo Edificio 6, via Cintia,
80126 Napoli, Italy}
\affiliation{Department of Physics, Faculty of Science,  The Maharaja Sayajirao University of Baroda, Vadodara 390002, Gujarat,  India}

\date{\today}

\begin{abstract}
In 2020, the LHCb collaboration reported the exclusive branching fractions for the channels $B_s^0 \to D_s^{(*)-}\mu^+\nu_\mu$ for the very first time.  In view of these observations,  we have recently reported the form factors and branching fraction computations for these channels employing the covariant confined quark model.  As different other channels corresponding to $b \to c \ell \nu_\ell$ have provided the hint for New Physics,  the analysis of observables such as forward-backward asymmetry,  longitudinal and transverse polarizations across the lepton flavours can serve as one of the important probes for the search for possible New Physics.  In present work,  we compute these observables for all the lepton flavours and compare our predictions with the other theoretical approaches.
\end{abstract}

\maketitle
\section{Introduction}
\label{sec:introduction}
For last many years, $b \to c \ell \nu_\ell$ has served as a very precise probe for the search of new physics.  As both the quarks involved in this transition are heavy,  it has great phenomenological implications within and beyond the standard model.  Experimentally, precise results are available for the channels $B \to D^{(*)}\ell\nu_\ell$ through different facilities such as BABAR, BELLE and LHCb collaborations.  Precise lattice results are also available corresponding to these transitions \cite{MILC:2015uhg,Na:2015kha}.   
Different heavy flavor anomalies corresponding to these transitions are reported in Ref.  \cite{Li:2018lxi,Albrecht:2021tul,London:2021lfn} and references therein.
Similarly,  $B_s \to D_s^{(*)}\ell \nu_\ell$ can also serve as a prominent channel for understanding the heavy flavor dynamics and possibly the anomalies.
On experimental side, LHCb have reported the branching fractions for the channels $B_s^0 \to D_s^{(*)-}\mu \nu_\mu$	 for the very first time \cite{Aaij:2020hsi} and determined the ratio of the branching fractions $B_s^0 \to D_s^- \mu^+ \nu_\mu$ to $B_s^0 \to D_s^{*-} \mu^+ \nu_\mu$.  Additionally, they also determined the ratios of these branching fractions relative to $B \to D$ namely
\begin{eqnarray}
\frac{\mathcal{B} (B_s^0  \to  D_s^- \mu^+ \nu_\mu)}{\mathcal{B}(B^0 \to D^- \mu^+ \nu_\mu)} &=&   
 1.09 \pm 0.05_{\mathrm{stat}} \pm 0.06_{\mathrm{syst}} \pm 0.05_{\mathrm{ext}} \nn
\frac{\mathcal{B}(B_s^0 \to D_s^{*-} \mu^+ \nu_\mu)}{\mathcal{B}(B^0 \to D^{*-} \mu^+ \nu_\mu)} &=&   1.06 \pm 0.05_{\mathrm{stat}} \pm 0.07_{\mathrm{syst}} \pm 0.05_{\mathrm{ext}}. \nonumber
\end{eqnarray}
The ratios of the decay widths from tau mode to electron mode for $D_s^{(*)}$ ($R(D_s)$ and $R(D_s^*)$) are yet to be measured experimentally so far.  However, lattice results are available for these ratio in the Refs.  \cite{McLean:2019qcx,Harrison:2021tol}.
New Physics studies have also been reported using these lattice form factors in Ref. \cite{Penalva:2023snz}.
Recently,  the ratios $R(D_s^{(*)})$ have been computed using unitarity and lattice QCD approach \cite{Martinelli:2022xir} where the transition form factors are computed in the entire momentum transfer range using the dispersive matrix approach.
The transition form factors, branching fractions and ratios $R(D_s^{(*)})$ are also computed using three point QCD sum rules \cite{Azizi:2008tt,Azizi:2008vt} and also using light cone QCD sum rules in the framework of heavy quark effective theory \cite{Bordone:2019guc,Zhang:2022opp}.
Transition form factors, branching fractions and other observables have also been computed recently within the framework of relativistic quark model (RQM) based on the quasi-potential approach in QCD \cite{Faustov:2012mt,Faustov:2022ybm}.
The transition form factors are also computed in both space- and time-like momentum transfer range within the constituent quark model framework \cite{Heger:2021gxt},employing the  next-to-leading-order QCD corrections \cite{Cui:2023bzr}  and also  employing light front quark model (LFQM) \cite{Verma:2011yw}.
Heavy to heavy and heavy to light semileptonic transitions form factors are also computed employing the symmetry preserving vector-vector contact interactions (SCI) \cite{Xing:2022sor}.
The detailed description of transition form factors and branching fractions are available in literature, however the detailed analysis of other physical observables such as forward backward asymmetry, different polarizations observables are yet to be explored.  As there is no indication of these observables from experimental side,  they may also serve as crucial probes for search for the Physics beyond the standard model.
In theoretical side, there are very few references in which these observables are studied, few of them are lattice results \cite{Harrison:2021tol},  RQM \cite{Faustov:2012mt} and light cone QCD sum rules \cite{Bordone:2019guc}.  
These observables are also studied considering the new physics scenario \cite{Dutta:2018jxz,Das:2019cpt,Das:2021lws}.

Very recently,  we have studied the transition form factors, branching fractions of these channels  namely $B_s \to D_s^{(*)}\ell \nu_\ell$ with $\ell = e,  \mu$ and $\tau$ within the quark model framework \cite{Soni:2021fky}. In this work, we have reported detailed analysis of the branching fractions with the recent LHCb data and it is observed that our results are in the excellent agreement with them along with lattice and other theoretical models.
In this paper, we provide much detailed description of the form factors with comparison to the lattice and other theoretical approaches.  We also provide the detailed plots for the different physical observables such as forward-backward asymmetries,  longitudinal and transverse polarization,  lepton and hadron side convexity parameter and so on.  Together with \cite{Soni:2021fky},  this work will provide complete description of the dynamics of semileptonic $B_s \to D_s^{(*)}$ decays.

Rest of the paper is organised in the following way.
After the brief introduction and some very recent literature survey,  for computation of the transition form factors, we provide very short description to our theoretical model that is covariant confined quark model in Sec.  \ref{sec:model} and provide the form factors in the double pole approximation. We also compare our results with  other theoretical approaches. Next in sec \ref{sec:observables}, we give the relations for the computation of the different physical observables.  We also provide the plots of these observables as well as the expectation values of these observables. In sec \ref{sec:result}, we discuss about all the results obtained for the semileptonic decay for the channels $B_s^0 \to D_s^{(*)-}\ell^+ \nu_\ell$ for $\ell = e, ~\mu,~\tau$. Finally, we conclude the present work in sec \ref{sec:conclusion}.
\section{Form factors}
\label{sec:model}
\begin{table*}[!ht]
\caption{Form factors and double pole parameters appeared in Eq. \ref{eq:form_factor_general} \cite{Soni:2021fky}.}
\begin{tabular*}{\textwidth}{@{\extracolsep{\fill}}cccccccc@{}}
\hline
$F$ & $F(0)$ & $a$ & $b$ & $F$ & $F(0)$ & $a$ & $b$\\
\hline
$F_+^{B_s \to D_s}$ 		& $0.770 \pm 0.066$ & 0.837 & 0.077 & $F_-^{B_s \to D_s}$ &  $-0.355 \pm 0.029$ & 0.855 & 0.083 \\
$A_+^{B_s \to D_s^*}$ 	& $0.630 \pm 0.025$ & 0.972 & 0.092 & $A_-^{B_s \to D_s^*}$ & $-0.756 \pm 0.031$ & 1.001 & 0.116\\
$A_0^{B_s \to D_s^*}$ 	& $1.564 \pm 0.065$ & 0.442 & $-$0.178 & $V^{B_s \to D_s^*}$ & $0.743 \pm 0.030$ & 1.010 & 0.118\\
\hline
\label{tab:form_factor}
\end{tabular*}
\end{table*}
\begin{figure*}[t]
\includegraphics[width=0.45\textwidth]{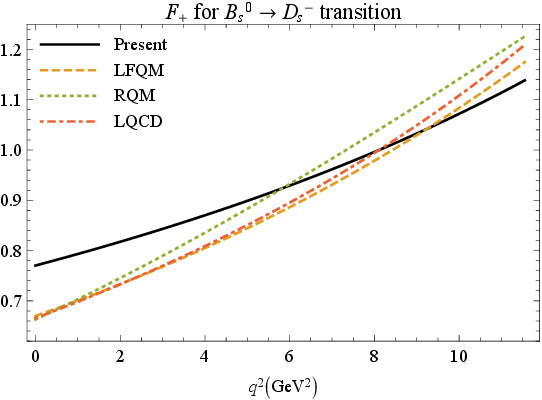}
\hfill\includegraphics[width=0.45\textwidth]{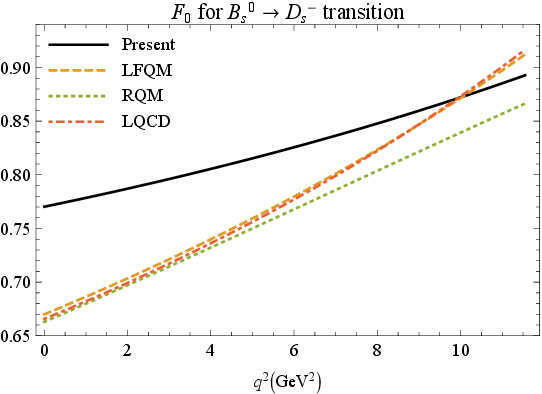}\\
\includegraphics[width=0.45\textwidth]{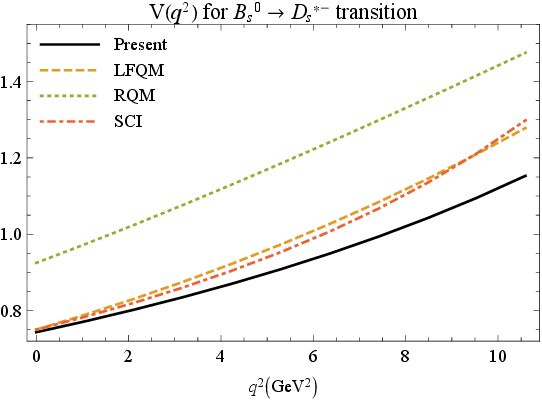}
\hfill\includegraphics[width=0.45\textwidth]{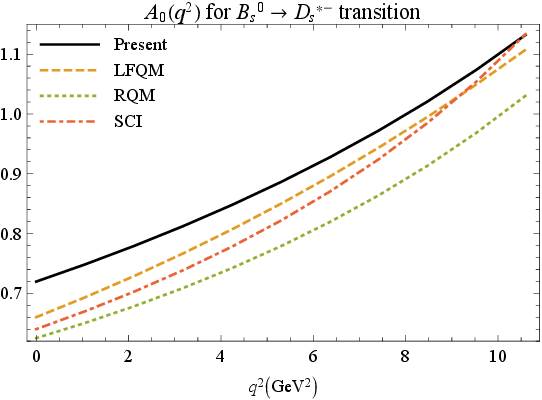}\\
\includegraphics[width=0.45\textwidth]{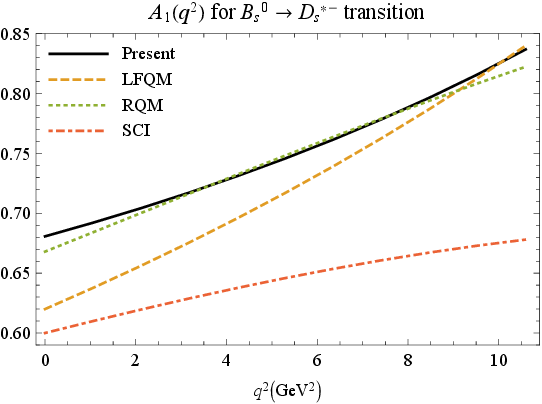}
\hfill\includegraphics[width=0.45\textwidth]{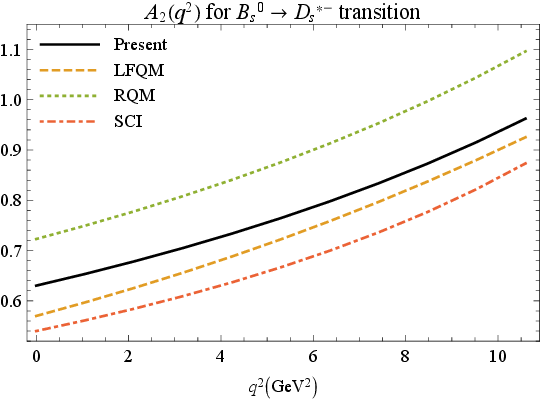}\\
\caption{Form factors in comparison with LFQM \cite{Verma:2011yw},  RQM \cite{Faustov:2012mt},  LQCD \cite{McLean:2019qcx} and SCI \cite{Xing:2022sor}.}
\label{fig:form_factor_BsDs}
\end{figure*}
Within the framework of standard model,  the matrix element for the semileptonic transition can be in the general form as
\begin{eqnarray}
\mathcal{M} (M_1 \to M_2^{(*)} \ell^+ \nu_\ell) = \frac{G_F}{\sqrt{2}} V_{\mathrm{CKM}} \langle M_2^{(*)} | \bar{q_2} O^\mu q_1|M_1 \rangle [\ell^+ O_\mu \nu_\ell],
\end{eqnarray}
This transition matrix element can also be parametrized in terms of form factors as
\begin{eqnarray}
\langle M_2 (p_2) &|& \bar{q_2} O^\mu q_1 | M_1 (p_1) \rangle  =  F_+ (q^2) P^\mu + F_- (q^2) q^\mu , \nn
\langle M_2^* (p_2, \epsilon_\nu) &|& \bar{q_2} O^\mu q_1 | M_1 (p_1) \rangle  =  \frac{\epsilon_{\nu}^{\dag}}{m_{M_1} + m_{M_2}} \left[ -g^{\mu\nu} P\cdot q A_0(q^2)  \right. \cr && \left. + P^{\mu} P^{\nu}  A_+(q^2) + q^{\mu} P^{\nu}  A_-(q^2) +  i\varepsilon^{\mu\nu\alpha\beta} P_{\alpha} q_{\beta}V(q^2) \right ],
\label{eq:form_factor_general}
\end{eqnarray}
In the above equations, $G_F$ is the Fermi coupling constant and $O^\mu = \gamma_\mu (1 - \gamma_5)$ is the weak Dirac matrix.  Also $P = p_1 + p_2$,  $q = p_1 - p_2$ and $\epsilon_{\nu}$ to be the polarization vector.  Further the on-shell conditions demands $p_1^2 = m_{M_1}^2$ and $p_2^2 = m_{M_2}^2$.
The form factors appearing in these equations are computed employing the effective field theoretical approach of covariant confined quark model (CCQM) originally developed by  M. A. Ivanov and G. V. Efimov \cite{Efimov:1988yd,Efimov:1993,Ivanov:1999ic,Branz:2009cd,Ivanov:2011aa,Gutsche:2012ze,Ivanov:2019nqd}.
The hadronic interaction Lagrangian can be written as
\begin{eqnarray}
\mathcal{L}_{\mathrm{int}}  &=& g_M M(x) \int dx_1\int dx_2 F_M(x;x_1,x_2) \bar{q}_2(x_2) \Gamma_M q_1(x_1) + \mathrm{H.c}.
\label{eq:lagrangian}
\end{eqnarray}
Above Lagrangian describes the interaction of meson with constituents.  Here the Dirac matrix $\Gamma$ is the gamma matrix according the meson.
$F_M(x;x_1,x_2)$ is the vertex function which essentially describes the distribution of quarks inside the hadron. In CCQM,  for computation of different observables, we consider the Gaussian form of the vertex function. It is important to note here that there are other forms of the vertex function also, but it is observed that the physical observables are not dependent on the detailed structure of the vertex functions.  Further,  Gaussian form also makes the analytical work more convenient \cite{Faessler:2003yf, Ivanov:1997ug}.
In Eq. \ref{eq:lagrangian},  $g_M$ is the coupling constant of meson computed employing the Compositeness conditions \cite{Salam:1962,Weinberg:1962}.  The computation of $g_M$ include the renormalization of self energy diagram.  This condition essentially confirms that the final mesonic state does not contain any free quark and all the quarks are confined within the hadron.
Further, the matrix element of the meson mass operator and semileptonic transition form factors are described by the convolution of the Feynman propagators and Gaussian vertex functions.
In order to have the loop integration in more efficient way, we use the Fock Schwinger representation of Feynman propagator.
Finally, to overcome any ultraviolet divergences appearing in the diagrams, we introduce the infrared cutoff parameter $\lambda = 0.181$ GeV.  Note here that we take $\lambda$ to be same for all the physical processes studied using the CCQM.

There are only two other model parameters namely quark mass and size parameter which are fixed by fitting with some basic properties such as leptonic decay widths with the available experimental data or lattice simulations.
The parametrization is also achieved in such a way that the deviation in the decay widths computation remains minimum.
The uncertainties in the fitted size parameters are determined from the differences in the predicted and experimental data.  It is worth interesting to note here that the observed uncertainness are found to be less than 5\% for all the mesonic case.
These uncertainties are then transported in the computation of all the observables such as form factors and then branching fractions and other physical observables.  It is observed that the uncertainties are less than 5\% at the $q^2 = 0$ and less than 10\% for $q^2 = q^2_{\mathrm{max}}$.
The propagation of uncertainty in any observables can be computed using the most general technique.  For instance, the uncertainty propagation in the forward backward asymmetry for the transition corresponding to the channel $B_s^0 \to D_s^-$ can be written as \cite{Soni:2020sgn}
\begin{eqnarray}
\Delta (A_{\mathrm{FB}}(q^2)) = \sqrt{\sum_{i = +, -} \left(\frac{\partial (A_{\mathrm{FB}}(q^2)) }{\partial F_i} \Delta F_i \right)^2}.
\end{eqnarray}
Here $\Delta F_i$ is the uncertainty in the form factor. 
For all the finer details regarding the computational techniques, we suggests the readers to refer Refs. \cite{Branz:2009cd,Lyubovitskij:2003pn}.

We also present the form factors Eq. (\ref{eq:form_factor_general}) in the double pole approximation form.  The relation reads
\begin{eqnarray}
F(q^2) = \frac{F(0)}{1 - a \left(\frac{q^2}{m_{M_1}^2}\right) + b \left(\frac{q^2}{m_{M_1}^2}\right)^2}.
\label{eq:double_pole}
\end{eqnarray}
The parameters $a$ and $b$ and the form factor $F(0)$ are listed in Tab. \ref{tab:form_factor}.
We also transform of our form factors with the Bauer-Stech-Wirbel form factors \cite{Wirbel:1985ji} so that we can have the brief comparison of our form factors in the entire $q^2$ range with different other theoretical approaches such as light front quark model \cite{Verma:2011yw}, relativistic quark model \cite{Faustov:2022ybm} and lattice simulations \cite{McLean:2019qcx}.
The transformed (primed) form factors reads
\begin{eqnarray}
F_0^\prime &=& F_+ + \frac{q^2}{m_{M_1}^2 - m_{M_2}^2} F_-,  \ \ \ \ \ \ F_+^\prime = F_+ \nn
A_0^\prime &=& \frac{m_{M_1} - m_{M_2}}{2m_{M_2}} \left(A_0 - A_+ - \frac{q^2}{m_{M_1}^2 - m_{M_2}^2} A_-\right),\\
A_1^\prime &=& \frac{m_{M_1} - m_{M_2}}{m_{M_1} + m_{M_2}} A_0,  \ \ \ \ \ \ A_2^\prime = A_+, \ \ \ \ \ \ V^\prime = V. \nonumber
\end{eqnarray}
Note that in order to avoid confusion, we will not use prime from now on.
In Fig. \ref{fig:form_factor_BsDs}, we compare transition form factors $F_+(q^2)$ and $F_0(q^2)$ with light front quark model, relativistic quark model and also the lattice simulations.

CCQM is a very versatile model and can be employed for studying not only mesons and baryons, but  also mutiquark states as well.
In last few years,  we have employed CCQM for studying the semileptonic decays of charmed mesons \cite{Soni:2017eug,Soni:2018adu,Ivanov:2019nqd,Soni:2020sgn,Soni:2019qjs,Soni:2019huk,Soni:2021rem},  bottom-strange mesons \cite{Soni:2021fky} and rare semileptonic decays of bottom mesons \cite{Soni:2020bvu,Soni:2021umv,Issadykov:2022gtr}.
Ivanov et. al., have also utilised CCQM for computation of various decay properties of $B$, $B_s$ and $B_c$ mesons in Ref. \cite{Ivanov:2000aj,Faessler:2002ut,Ivanov:2005fd,Ivanov:2006ni,Ivanov:2011aa,Dubnicka:2017job,Issadykov:2018myx,Dubnicka:2018gqg,Dubnicka:2022gsi,Ivanov:2022nnq}.
\begin{table*}[!ht]
\caption{Averages of different physical observables for $B_s^0 \to D_s^- \ell^+ \nu_\ell$ channel.}
\begin{tabular*}{\textwidth}{@{\extracolsep{\fill}}ccccccc@{}}
\hline
Observable	&	$-\langle A_{FB}^e \rangle  \times 10^{-6}$ 	&	$-\langle A_{FB}^\mu \rangle$	&	$-\langle A_{FB}^\tau   \rangle$	&	$-\langle C_{F}^e \rangle$	&	$-\langle C_{F}^\mu \rangle$	&	$-\langle C_{F}^\tau \rangle$	\\
\hline													
Present 	&	$1.156 \pm 0.295$	&	$0.015 \pm 0.004$	&	$0.362 \pm 0.113$	&	$1.500 \pm 0.395$	&	$1.455 \pm 0.384$	&	$0.260 \pm 0.079$	\\
RQM \cite{Faustov:2022ybm}	&	$0.97$	&	$0.013$	&	$0.36$	&	$1.5$	&	$1.46$	&	$0.3$	\\
\hline\hline													
Observable 	& 	$\langle P_L^e \rangle$	& 	$\langle P_L^\mu \rangle$	&	$-\langle P_L^\tau\rangle$	&	$-\langle P_T^e \rangle  \times 10^{-3}$	& 	$-\langle P_T^\mu \rangle$	&	$-\langle P_T^\tau\rangle$	\\
\hline													
Present 	& 	$1.000 \pm 0.263$	& 	$0.958 \pm 0.253$	&	$0.337 \pm 0.119$	&	$1.115 \pm 0.294$	&	$0.205 \pm 0.054$	&	$0.839 \pm 0.264$	\\
RQM \cite{Faustov:2022ybm}	& 	1.000	& 	0.960	&	$0.27$	&	$1.02$	&	$0.19$	&	$0.85$	\\
\hline
\label{tab:Obs_Bs_Ds}
\end{tabular*}
\end{table*}
\begin{table*}[!ht]
\caption{Averages of different physical observables for $B_s^0 \to D_s^{-*} \ell^+ \nu_\ell$ channel.}
\begin{tabular*}{\textwidth}{@{\extracolsep{\fill}}ccccccc@{}}
\hline
Observable	&	$-\langle A_{FB}^e \rangle$ 	&	$-\langle A_{FB}^\mu \rangle$	&	$-\langle A_{FB}^\tau   \rangle$	&	$-\langle C_{F}^e \rangle$	&	$-\langle C_{F}^\mu \rangle$	&	$-\langle C_{F}^\tau \rangle$	\\
\hline													
Present 	&	$0.195 \pm 0.024$	&	$0.201 \pm 0.025$	&	$0.284 \pm 0.037$	&	$0.452 \pm 0.089$	&	$0.436 \pm 0.086$	&	$0.060 \pm 0.012$	\\
RQM \cite{Faustov:2022ybm}	&	$0.26$	&	$0.27$	&	$0.32$	&	$0.35$	&	$0.33$	&	$0.040$	\\
\hline\hline												
Observable 	& 	$\langle P_L^e \rangle$	& 	$\langle P_L^\mu \rangle$	&	$\langle P_L^\tau\rangle$	&	$-\langle P_T^e \rangle \times 10^{-3}$	& 	$-\langle P_T^\mu \rangle$	&	$-\langle P_T^\tau\rangle$	\\
\hline													
Present 	& 	$1.000 \pm 0.148$	& $	0.985 \pm 0.145$	&	$0.504 \pm 0.067$	&	$0.322 \pm 0.063$	&	$0.058 \pm 0.011$	&	$0.126 \pm 0.034$	\\
RQM \cite{Faustov:2022ybm}	& 	1.000	& 	0.990	&	0.530	&	$0.23$	&	$0.040$	&	$0.035$	\\
\hline\hline												
Observable 	& 	$\langle F_L^e \rangle$	& 	$\langle F_L^\mu \rangle$	& 	$\langle F_L^\tau \rangle$	\\
	\hline													
Present 	& 	$0.534 \pm 0.088$	& $0.534 \pm 0.088$	& $0.458 \pm 0.070$	\\						
RQM \cite{Faustov:2022ybm}	& 	0.49	& 	0.49	& 	0.42	\\	
LQCD \cite{Harrison:2021tol}		& -- & -- & 0.440 (16)\\			
\hline
\label{tab:Obs_Bs_Dsv}
\end{tabular*}
\end{table*}
\section{Different physical observables}
\label{sec:observables}
After computation of transition form factors, we compute various other physical observables such as forward-backward asymmetry,  lepton side convexity parameter, longitudinal and transverse polarization.
In Ref. \cite{Soni:2021fky}, we have computed the semileptonic branching fractions and made a detailed comparison with the LHCb data and here in present work, we compute some other physical observables that are yet to be identified by the experimental facilities worldwide.
These observables are forward backward asymmetry, convexity parameter, longitudinal and transverse polarization of charged leptons, and longitudinal polarization fractions of daughter vector meson.
In semileptonic decays, these observables are dependent on the lepton masses and therefore these observables are very important probes to understand the effects of lepton masses. 
In present work, we use the same notations used for studying the semileptonic $D$ and $D_s$ decays \cite{Ivanov:2019nqd}.
We compute the following physical observable in the entire $q^2$ range and their relations are defined as \cite{Ganbold:2014pua,Gutsche:2015mxa,Zhang:2020dla,Faustov:2022ybm}:
\begin{enumerate}
\item Forward-backward asymmetry:
\begin{eqnarray}
A_{\mathrm{FB}} (q^2) = \frac34 \frac{\mathcal{H}_P - 2 \frac{m_\ell^2}{q^2} \mathcal{H}_{SL}}{\mathcal{H}_{tot}}
\label{eq:asymmetry}
\end{eqnarray}
\item Lepton side convexity parameter:
\begin{eqnarray}
C_F^\ell = \frac34 \left(1 - \frac{m_\ell^2}{q^2}\right) \frac{\mathcal{H}_U - 2 \mathcal{H}_L}{\mathcal{H}_{tot}}
\label{eq:convex}
\end{eqnarray}
\item Longitudinal polarization of charged leptons:
\begin{eqnarray}
P_L^\ell = \frac{(\mathcal{H}_U + \mathcal{H}_L) \left(1 - \frac{m_\ell^2}{2q^2}\right) - \frac{3 m_\ell^2}{2q^2}\mathcal{H}_s}{\mathcal{H}_{tot}}
\label{eq:polarization_l}
\end{eqnarray}
\item Transverse polarization of charged leptons:
\begin{eqnarray}
P_T^\ell = - \frac{3 \pi m_\ell}{8 \sqrt{q^2}}  \frac{\mathcal{H}_P + 2 \mathcal{H}_{SL}}{\mathcal{H}_{tot}}
\end{eqnarray}
\label{eq:polarization_t}
\item Longitudinal polarization fraction for the final vector meson:
\begin{eqnarray}
F_L(q^2) = \frac{\mathcal{H}_L \left(1 + \frac{m_\ell^2}{2q^2}\right) + \frac{3 m_\ell^2}{2 q^2} \mathcal{H}_S}{\mathcal{H}_{tot}}
\label{eq:Fl}
\end{eqnarray}
\end{enumerate}
Using the above Eq. (\ref{eq:Fl}), one can also compute the transverse polarization vector of the final vector mesons via relation $F_T = 1 - F_L$.
In these relations,  $\mathcal{H}'s$ are the bilinear combinations of the helicity components of the hadronic tensors related to the helicity amplitudes ($H$) as \cite{Ganbold:2014pua,Gutsche:2015mxa,Zhang:2020dla,Faustov:2022ybm},
\begin{eqnarray}
\mathcal{H}_U &=&  |H_+|^2 + |H_-|^2,  \ \ \ \ \ \ \mathcal{H}_P = |H_+|^2 - |H_-|^2, \nn
\mathcal{H}_L &=& |H_0|^2,  \ \ \ \ \ \ \mathcal{H}_S = |H_t|^2,  \nn 
\mathcal{H}_{SL} &=& Re (H_0 H_t^\dag).
\end{eqnarray}
Here the helicity amplitudes are related to the transition form factors computed using CCQM. The relations read
\begin{enumerate}
\item For $B_s^0 \to D_s^-$ transitions
\begin{eqnarray}
H_t &=& \frac{1}{\sqrt{q^2}} (Pq F_+ + q^2 F_-),\nn
H_\pm &=& 0 \ \ \ {\textrm and} \ \ \ H_0 = \frac{2 m_{M_1} |\bf{p_2}|}{\sqrt{q^2}} F_+,
\end{eqnarray}
\item For $B_s^0 \to D_s^{-*}$ transitions
\begin{eqnarray}
H_t  &=& \frac{1}{m_{M_1}+m_{M_2}} \frac{m_{M_1} |{\bf p_2}|}{m_{M_2}\sqrt{q^2}}  \left((m_{M_1}^2 - m_{M_2}^2) (A_+ - A_-) + q^2 A_-\right)\nn
H_\pm &=& \frac{1}{m_{M_1}+m_{M_2}} (-(m_{M_1}^2 - m_{M_2}^2)A_0 \pm 2 m_{M_1} |{\bf p_2}| V) \nn
H_0 &=& \frac{1}{m_{M_1}+m_{M_2}} \frac{1}{2m_{M_2}\sqrt{q^2}} (-(m_{M_1}^2 - m_{M_2}^2)(m_{M_1}^2 - m_{M_2}^2-q^2)A_0 + 4m_{M_1}^2 |{\bf p_2}|^2A_+).\nonumber
\end{eqnarray}
\end{enumerate}
Here the form factors are defined in Eq. (\ref{fig:form_factor_BsDs}).
Also the $|{\bf p_2}| = \lambda^{1/2} (m_{M_1}^2,m_{M_2}^2,q^2)/2m_{M_1}$ is the momentum of the $D_s^{(*)-}$ meson in the rest frame of $B_s^0$ mesons and $\lambda$ is the K\"allen function. Further, in all these equations $M_1$ is the parent meson ($B_s^0$) and $M_2$ is the daughter mesons ($D_s^{(*)-}$).
Using these relations we compute the physical observables and listed in Tab. \ref{tab:Obs_Bs_Ds} and \ref{tab:Obs_Bs_Dsv}.
We also compare our findings with the relativistic quark model predictions \cite{Faustov:2022ybm}.
Note here that in order to compute the expectation values of these observables, one has to multiply with the phase factor $|{\bf p_2}| (q^2 - m_\ell^2)q^2$ to the numerator and denominator and integrate separately.
The CCQM model parameters used for the present computations are quark masses $m_b = 5.05$ GeV,  $m_c = 1.672$ GeV, $m_s = 0.428$ GeV and meson size parameters $\Lambda_{B_s} = 2.05 \pm 0.014$ GeV, $\Lambda_{D_s} = 1.75 \pm 0.035$ GeV and $\Lambda_{D_s^*} =  1.56 \pm 0.014$ GeV \cite{Soni:2021fky}.

\section{Results and Discussion}
\label{sec:result}
Having determined the model parameters namely quark masses and size parameters, we compute the transition form factors as per Eq.  (\ref{eq:form_factor_general}) in the entire $q^2$ range.
In fig. \ref{fig:form_factor_BsDs} we plot the form factors and also compare our form factors with light front quark model \cite{Verma:2011yw}, relativistic quark model \cite{Faustov:2022ybm}, SCI \cite{Xing:2022sor} as well as with the lattice simulations \cite{McLean:2019qcx}.
For $B_s^0 \to D_s^-$ form factors, our results are very near to other theoretical approaches including lattice simulations. However, for $F_0(q^2 < 10 \mathrm{GeV}^2)$, our form factors are significantly higher than the other approaches. Our results for vector form factors are in very good agreement with the other approaches.

\begin{figure*}[htbp]
\includegraphics[width=0.45\textwidth]{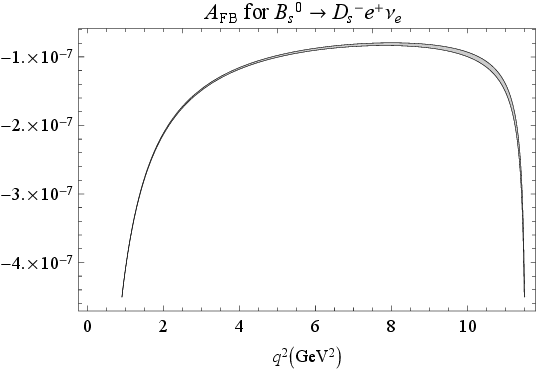}
\hfill\includegraphics[width=0.45\textwidth]{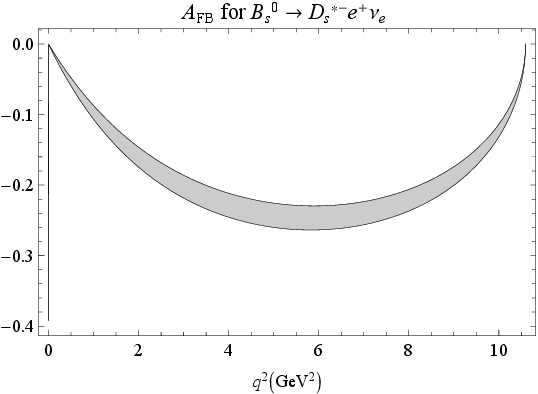}\\
\includegraphics[width=0.45\textwidth]{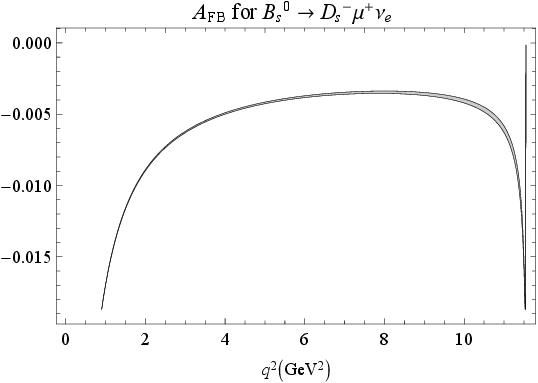}
\hfill\includegraphics[width=0.45\textwidth]{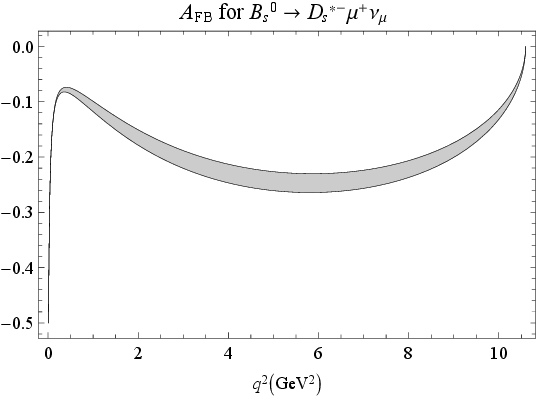}\\
\includegraphics[width=0.45\textwidth]{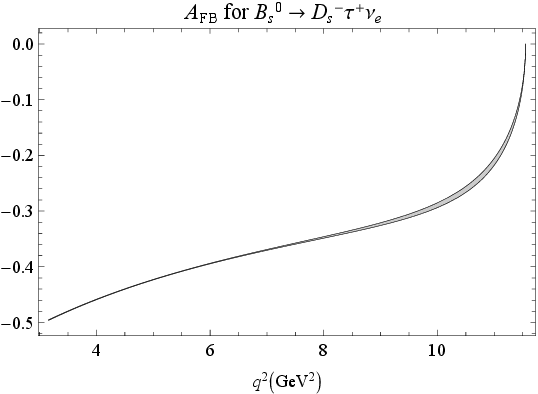}
\hfill\includegraphics[width=0.45\textwidth]{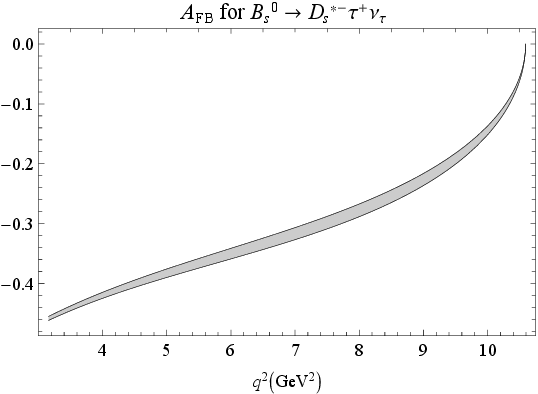}\\
\caption{Forward backward asymmetries for the decay channels $B_s^0 \to D_s^{(*)-}\ell^+\nu_\ell$.}
\label{fig:forward_backward_asymmetry}
\end{figure*}

\begin{figure*}[htbp]
\includegraphics[width=0.45\textwidth]{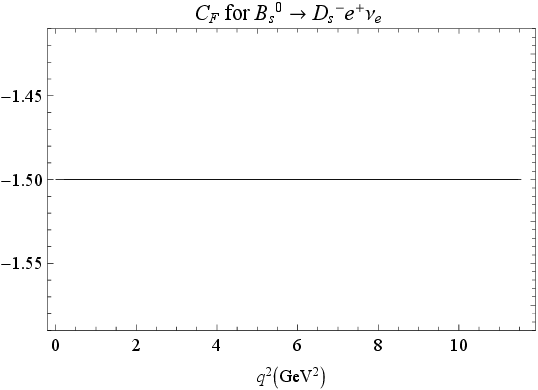}
\hfill\includegraphics[width=0.45\textwidth]{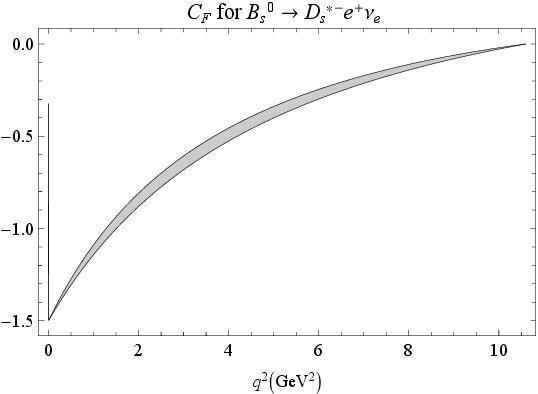}\\
\includegraphics[width=0.45\textwidth]{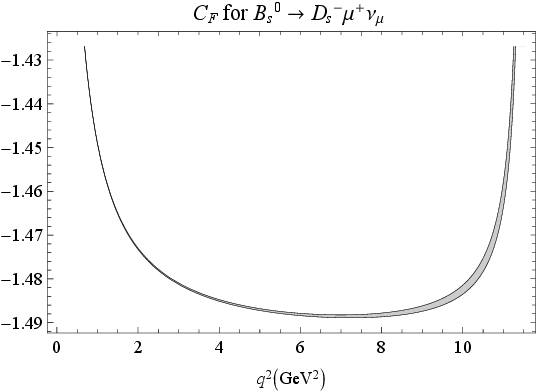}
\hfill\includegraphics[width=0.45\textwidth]{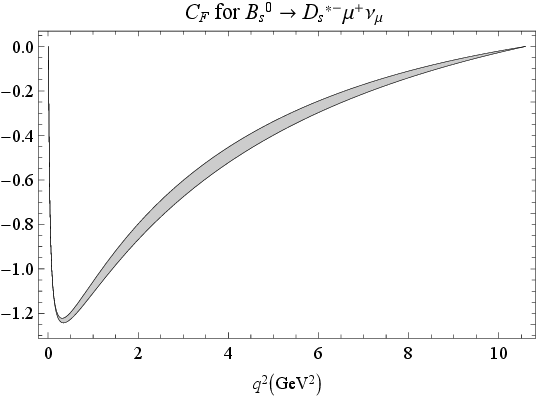}\\
\includegraphics[width=0.45\textwidth]{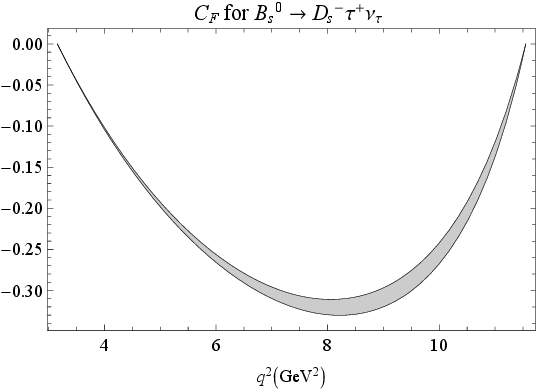}
\hfill\includegraphics[width=0.45\textwidth]{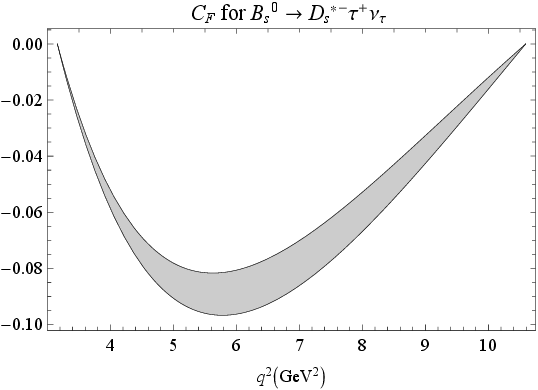}\\
\caption{Lepton side convexity parameter for the decay channels $B_s^0 \to D_s^{(*)-}\ell^+\nu_\ell$.}
\label{fig:convex}
\end{figure*}

\begin{figure*}[htbp]
\includegraphics[width=0.45\textwidth]{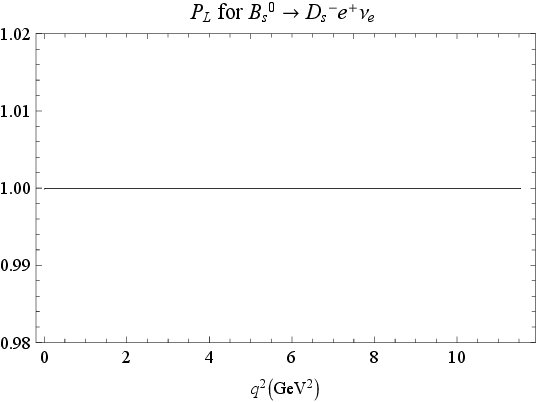}
\hfill\includegraphics[width=0.45\textwidth]{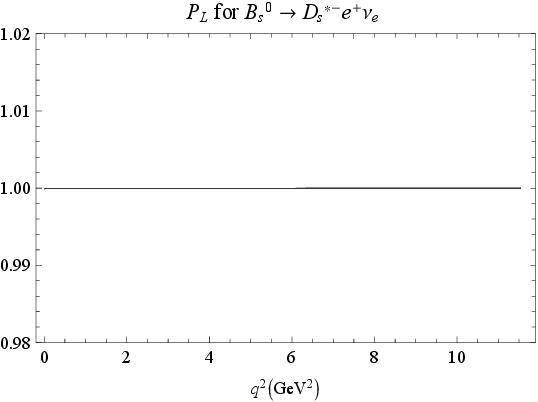}\\
\includegraphics[width=0.45\textwidth]{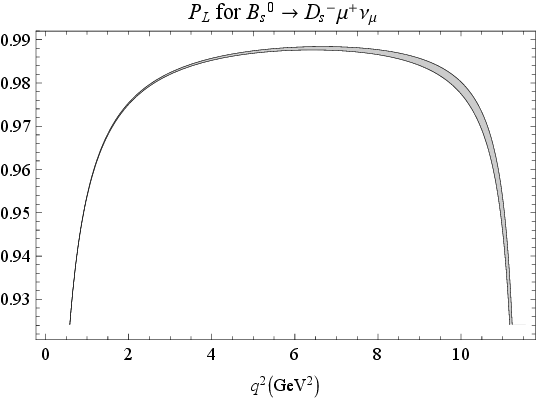}
\hfill\includegraphics[width=0.45\textwidth]{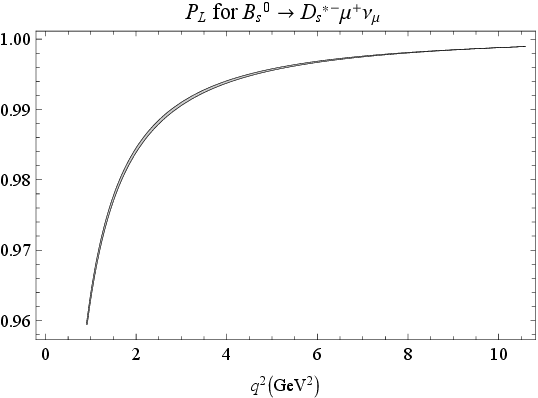}\\
\includegraphics[width=0.45\textwidth]{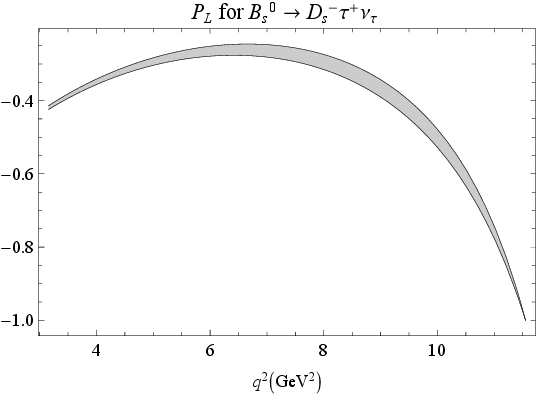}
\hfill\includegraphics[width=0.45\textwidth]{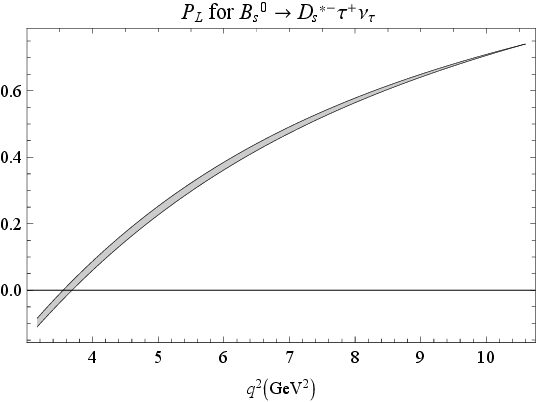}\\
\caption{Longitudinal polarization of charged leptons for the decay channels $B_s^0 \to D_s^{(*)-}\ell^+\nu_\ell$.}
\label{fig:longitudinal}
\end{figure*}

\begin{figure*}[htbp]
\includegraphics[width=0.45\textwidth]{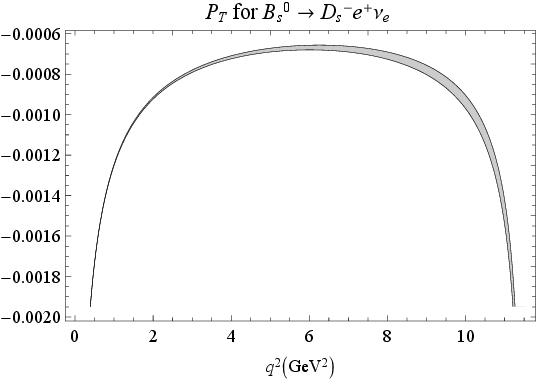}
\hfill\includegraphics[width=0.45\textwidth]{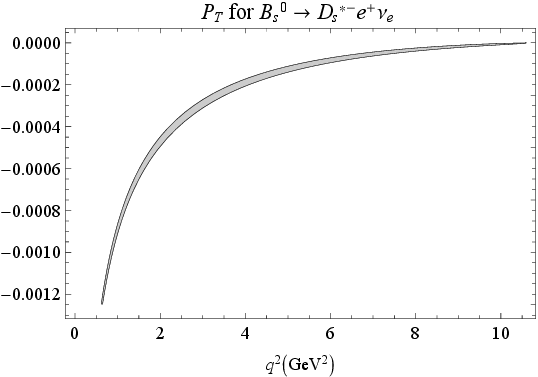}\\
\includegraphics[width=0.45\textwidth]{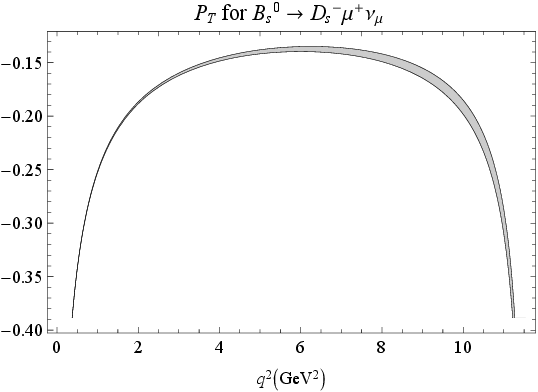}
\hfill\includegraphics[width=0.45\textwidth]{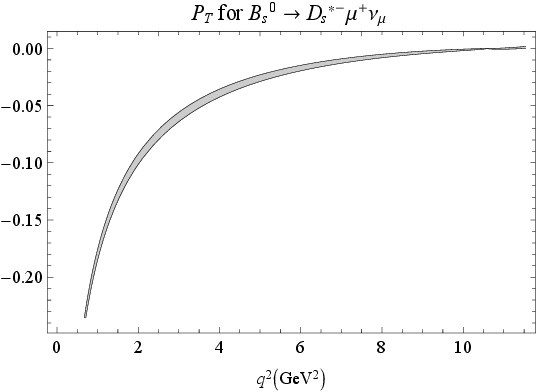}\\
\includegraphics[width=0.45\textwidth]{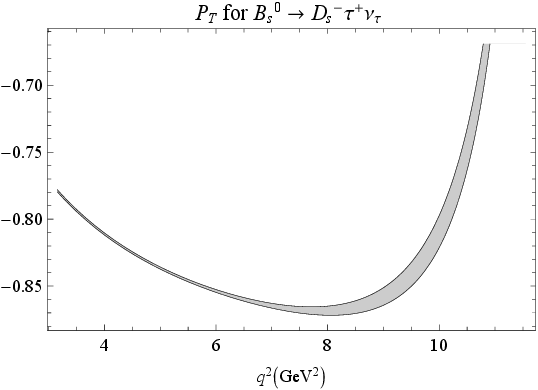}
\hfill\includegraphics[width=0.45\textwidth]{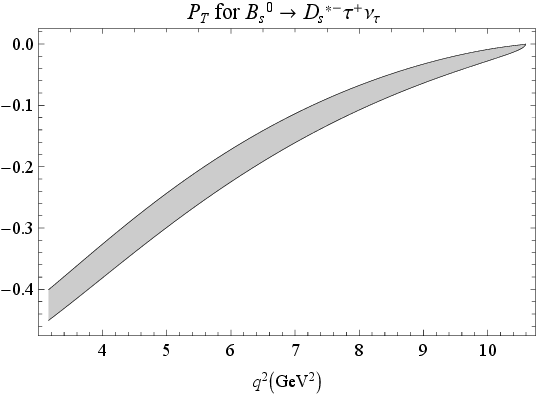}\\
\caption{Transverse polarization of charged leptons for the decay channels $B_s^0 \to D_s^{(*)-}\ell^+\nu_\ell$.}
\label{fig:transverse}
\end{figure*}

\begin{figure*}[htbp]
\includegraphics[width=0.45\textwidth]{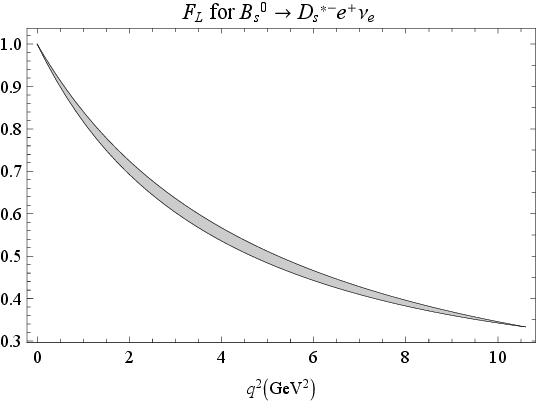}
\hfill\includegraphics[width=0.45\textwidth]{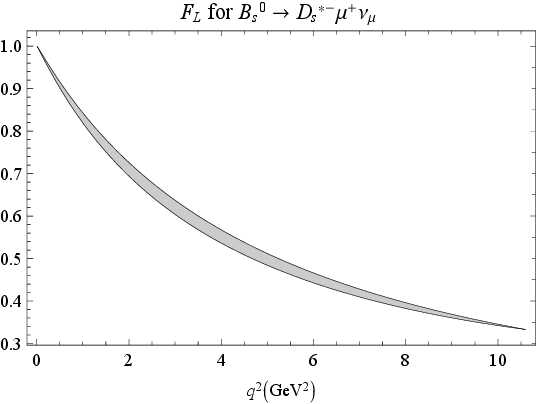}\\
\includegraphics[width=0.45\textwidth]{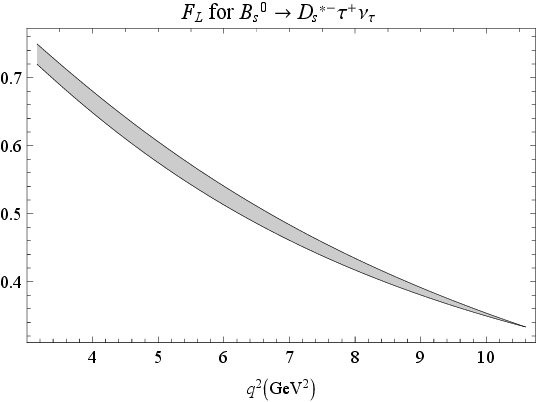}
\caption{Longitudinal polarization fraction for final state $D_s^{*-}$ meson.}
\label{fig:longitudinal_fraction}
\end{figure*}

Using the transition form factors we compute different physical observables using the relations Eq. (\ref{eq:asymmetry} - \ref{eq:Fl}) and their expectation values are listed in Tab. \ref{tab:Obs_Bs_Ds} and \ref{tab:Obs_Bs_Dsv}.
These observables are yet to be identified experimentally,  however,  for $B \to D^{(*)}$ transitions, these observables are studied extensively by the $B$ factories.  Therefore these observables are also expected for $B_s \to D_s^{(*)-}$ transitions from the experiments these channels are much similar to that of $B \to D^{(*)}$ except for the spectator quark.
We also compare our results with the predictions using the relativistic quark model by Faustov \textit{et. al}., \cite{Faustov:2022ybm} and it is observed that our results are in excellent agreement with them for all the observables.  This agreement is expected as our form factors are also nearly matching well.
In Figs. \ref{fig:forward_backward_asymmetry} - \ref{fig:longitudinal_fraction}, we also plot all these observables in the whole kinematical range of momentum transfer squared along with the spread of uncertainty. 
It is observed that the lepton side convexity parameter for the channel $B_s^0 \to D_s^- e^+\nu_e$ and longitudinal polarization for the channels $B_s^0 \to D_s^{(*-)} e^+\nu_e$ are found to be constant throughout the whole $q^2$ range and also their spread in the uncertainty is also found to be very small and constant.
Recently, HPQCD collaboration \cite{Harrison:2021tol} have also provided the results on longitudinal polarization fraction for the $D_s^{(*)-}$ meson and from Tab. \ref{tab:Obs_Bs_Dsv}, it is observed that our results are in excellent agreement with them. 
It is interesting to note here that the lepton side convexity parameter for the channel $B_s^0 \to D_s^- e^+ \nu_e$ and longitudinal polarization of charged leptons for the channel $B_s^0 \to D_s^{(*)-} e^+ \nu_e$ is found to be constant through out the $q^2$ range.  This nature is also observed in the RQM predictions \cite{Faustov:2022ybm}.

In our previous study \cite{Soni:2021fky}, we have  reported the semileptonic branching fractions and also provided  detailed comparison with the recent LHCb measurements as well as with the lattice simulation results. We determined the normalised decay rates for the channel $B_s^0 \to D_s^{*-}\mu^+ \nu_\mu$ in the recoil parameter bins and they are found to be in very good agreement with the LHCb as well as  lattice results.
We also determined the ratios $R(D_s) = 0.271 \pm 0.069$ and $R(D_s^{*-}) = 0.240 \pm 0.034$ and they are in very good agreement with the other theoretical approaches and lattice simulations within the uncertainties \cite{Soni:2021fky}.
These ratios are also in agreement with the other channels concerning the transition $b \to c \tau\nu_\tau/b \to c \mu\nu_\mu$ such as $B \to D^{(*)}\ell\nu_\ell$ \cite{Ivanov:2015tru} as well as $B_c \to (\eta_c,J/\psi)\ell\nu_\ell$ \cite{Issadykov:2017wlb} studied employing CCQM.
Further, the ratio $\mathcal{B} (B_s^0 \to D_s^- \mu^+\nu_\mu)/\mathcal{B} (B_s^0 \to D_s^{*-} \mu^+\nu_\mu) = 0.451 \pm 0.096$ is also found to be in excellent agreement with the LHCb data \cite{Soni:2021fky}.
Overall, all the results presented employing CCQM are in very good agreement with the available experimental data and lattice simulations. The same was also observed for studying the semileptonic decays of $D$ and $D_s$ mesons \cite{Soni:2017eug,Soni:2018adu,Ivanov:2019nqd,Soni:2020sgn}.

\section{Conclusion}
\label{sec:conclusion}
In this article, we have studied a very interesting channel corresponding to the quark level transition $b \to c \ell^+ \nu_\ell$.  $B \to D^{(*)} \ell \nu_\ell$ have gained lot of attention as some of its results deviate from the standard model predictions,  the another channel $B_s^0 \to D_s^{(*)} \ell^+\nu_\ell$ have started getting attention after the first observation from the LHCb collaboration.
Here, we compute the transition form factors for the channels $B_s \to D_s^-$ and $B_s^0 \to D_s^{-*}$ using the covariant confined quark model.  We also compare our form factors with different other theoretical predictions including lattice simulations.
We further compute different physical observables such as forward backward asymmetry, lepton side convexity parameter, longitudinal and transverse polarizations along with their expectation values.
Since these observables are yet to be examined from the experimental side, we compare only with the available theoretical prediction relativistic quark model.
In general,  all our results are in very good agreement with the available theoretical studies and lattice predictions.

\section*{ACKNOWLEDGEMENTS}
We would like to thank Prof. Mikhail A. Ivanov for useful discussions on some aspects of this work.

\section*{Data Availability Statement}
There are no data associated with the manuscript.

\clearpage
\bibliography{apssamp}

\end{document}